# Generation of spatiotemporal acoustic vortices with arbitrarily oriented orbital angular momentum

Shuai Liu [1], Hao Ge [1, *], Xiang-Yuan Xu [1], Yuan Sun [1], Xiao-Ping Liu [2], Ming-Hui Lu [1,3,4, †], and Yan-Feng Chen [1, 4, ‡]

[1] *National Laboratory of Solid State Microstructures & Department of Materials Science and Engineering, Nanjing University, Nanjing, Jiangsu 210093, China*

[2] *School of Physical Science and Technology, ShanghaiTech University, Shanghai 201210, China*

[3] *Jiangsu Key Laboratory of Artificial Functional Materials, Nanjing, Jiangsu 210093, China.*

[4] *Collaborative Innovation Center of Advanced Microstructures, Nanjing University, Nanjing 210093, China*

*Correspondence should be addressed to: haoge@nju.edu.cn, luminghui@nju.edu.cn or yfchen@nju.edu.cn*

**Abstract.** Despite extensive exploration of acoustic vortices carrying orbital angular momentum (OAM), the generation of acoustic vortices with OAM orientations beyond the conventional longitudinal direction remains largely unexplored. Spatiotemporal (ST) vortices, featuring spiral phase twisting in the ST domain and carrying transverse OAM, have recently attracted considerable interest in optics and acoustics. Here, we report the generation of three-dimensional (3D) ST acoustic vortices with arbitrarily oriented OAM, thereby opening up a new dimension in acoustic OAM control. By utilizing a two-dimensional (2D) acoustic phased array, we introduce two approaches to manipulate the orientation of OAM: through the direct rotation of vortices in 3D space and the intersection of vortices carrying distinct types of OAM. These methods enable unprecedented control over the orientation of acoustic OAM, providing a new degree of freedom in the manipulation of acoustic waves. The arbitrarily oriented OAM holds promise for enhancing acoustic communication by broadening capacity and enabling more complex particle manipulation techniques. Our work establishes a foundation for future explorations into the complex dynamics of novel structured acoustic fields in the ST domain.

*Introduction.*

Vortices are ubiquitous phenomena across various domains in physics. They are characterized by a doughnut-shaped intensity distribution and a spiral phase distribution, and have been extensively studied and generated in many wave fields, ranging from electromagnetic waves[1–5], acoustic waves[6–12] to electron waves[13] and neutron waves[14]. The vortex beams can carry longitudinal orbital angular momentum (OAM) aligned with their propagation direction, which provides an OAM spatial degree of freedom and has been widely utilized in applications such as high-capacity information transmission[15], super-resolution imaging[16], quantum key distribution[17], and particle manipulation[18,19].

Recently, there has been a growing interest in spatiotemporal (ST) vortices where the wave packets are structured in both space and time[20–30]. Differing from conventional monochromatic vortices with a spiral phase in the spatial domain, these vortices are polychromatic wave packets and exhibit a spiral phase in the ST domain. The energy circulation in the ST domain results in a transverse OAM perpendicular to the propagation direction. This transverse OAM provides a new degree of freedom which is expected to extend the functionalities of vortices, and leads to the discovery of novel physical phenomena, such as spin-orbit interaction between transverse spin and orbital angular momentum[24], transverse shifts and time delays of vortices at interfaces[31] and nonlinear optical harmonic generation with transverse OAM[32–34]. Generating such transverse OAM requires simultaneously spatial and temporal modulations of wave packets. In optics, ST vortices were generated experimentally using sophisticated optical systems containing components like spatial light modulators and lenses[35]. ST vortices can also be generated by designed structures such as photonic crystal slabs and resonant diffractive gratings. The transmission function of these structures exhibits nodal lines in the wavevector-frequency space[36–40]. Moreover, arbitrarily oriented optical vortices could be constructed[36,41–43], promising more controllable degree of freedoms and advanced applications.

Wave fields with OAM have also been widely investigated in the area of acoustics. Acoustic vortices can be generated by using a planar array of transducers[7,44] or passive metasurface structures[10]. Several acoustic OAM-based applications were proposed, such as acoustic trapping[45], rotation[46], levitation[47], and sound asymmetric transmission[48]. Acoustic communication by multiplexing OAM was also experimentally demonstrated, enabling high-speed data transmission for underwater applications[49]. Very recently, ST vortices have been explored in acoustics, bringing novel functionalities to acoustic OAM-based applications. The ST acoustic vortices are generated via a one-dimensional acoustic phased array[50] or a periodic meta-grating[51]. These vortices exist in the two-dimensional (2D) space and carry purely transverse OAM.

In this work, we advance our understanding of acoustic vortices by synthesizing three-dimensional (3D) ST acoustic vortices with OAM oriented in arbitrary directions, aiming for more comprehensive control over acoustic OAM. Compared to the 2D case, adding an additional dimension allows for the realization of more complex ST acoustic vortices. Unlike the methods used in optics, we decompose the ST vortices into a series of plane wave modes and generate them using an acoustic phased array. The 3D profile of ST acoustic vortices can be measured directly and precisely. We initially utilize a 2D phased array to generate 3D ST vortices carrying longitudinal and transverse OAM, respectively. To generate acoustic OAM with arbitrary orientations, we employ two distinct approaches. The first approach involves the direct rotation of wave packets carrying longitudinal or transverse OAM in 3D space, as depicted in Fig. 1(a). The orientation of OAM will change accordingly. The second approach is achieved through the intersection of vortices carrying longitudinal and transverse OAM, as shown in Fig. 1(b). By modifying the topological charges of these vortices, the orientation of the tilted OAM can be adjusted. The orientation of OAM can be regarded as a new degree of freedom, offering potential advantages in the field of acoustic communications. Furthermore, the tilted OAM can be employed for more sophisticated particle manipulation through the transfer of acoustic OAM to particles. Our work establishes a platform for investigating 3D ST acoustic vortices and may open new avenues for exploring novel structured acoustic fields in the ST domain.

## Generation of 3D ST vortices with longitudinal and transverse OAM.

Assuming that all wave packets propagate along the $z$-axis, we initially construct 3D ST acoustic wave packets carrying longitudinal and transverse OAM separately. To generate a 3D ST wave packet with longitudinal OAM, we introduce a spiral phase distribution on the $x-y$ domain of a 3D Gaussian wave packet. The acoustic pressure field can be expressed as follows:

$$p(x,y,z,t) \propto [x + i\,sgn(l)y]^{|l|} e^{-\frac{x^2+y^2+\zeta^2}{w^2}} e^{ik_0\zeta}, \qquad (1)$$

where the integer $l$ is the topological charge, $k_0$ is the central wavevector. The widths of the wave packet along $x$, $y$, and $z$ axes are identical and are defined by the parameter $w$. $\zeta = z - ct$ is the pulse-accompanying coordinate, where $c$ is the speed of acoustic wave. Similarly, the spiral phase can also be added on the $x-z$ domain (or the $x-t$ space-time domain), and the acoustic pressure field of the 3D Gaussian wave packet carrying transverse OAM can be expressed as follows:

$$p(x,y,z,t) \propto [\zeta + i\,sgn(l)x]^{|l|} e^{-\frac{x^2+y^2+\zeta^2}{w^2}} e^{ik_0\zeta}. \qquad (2)$$

The ST vortices are polychromatic wave packets that are constructed as a superposition of plane waves $P(\mathbf{k})e^{-i\omega(\mathbf{k})t}$, where $P(\mathbf{k})$ is the complex plane-wave Fourier amplitudes. The normalized integral values of OAM can be calculated using $P(\mathbf{k})$ as[52]:

$$\mathbf{L} = \frac{\int \omega^{-1} P^*(-i\mathbf{k} \times \nabla_{\mathbf{k}}) P d^3\mathbf{k}}{\int \omega^{-1} |P|^2 d^3\mathbf{k}}. \qquad (3)$$

It is worth noting that the OAM consists of both intrinsic and extrinsic components. The extrinsic component originates from the transverse shift of the pulse probability centroid $\mathbf{R}$, which is defined as[52]:

$$\mathbf{R} = \frac{\int \omega^{-1} P^* e^{i\omega t}(i\nabla_{\mathbf{k}}) P e^{-i\omega t} d^3\mathbf{k}}{\int \omega^{-1} |P|^2 d^3\mathbf{k}}. \qquad (4)$$

The extrinsic component satisfies $\mathbf{L}^{ext} = \mathbf{R} \times k_0$, and we can obtain the intrinsic component as:

$$\mathbf{L}^{int} = \mathbf{L} - \mathbf{L}^{ext}. \qquad (5)$$

The calculated $\mathbf{L}^{int}$ of the two wave packets defined by Eq. (1) and (2) are $(0,0,l)$ and $(0,l,0)$, which confirms that these two wave packets carry longitudinal and transverse OAM, respectively.

Our experimental setup is schematically shown in Figs. 1(c) and 1(d). The experiment is conducted in an anechoic chamber to mitigate the impact of reflections on the measurements. A 2D acoustic phased array consisting of 121 loudspeakers with a spacing of 5.5cm is utilized to generate 3D ST acoustic vortices. By performing a Fourier transform on the acoustic pressure field, we can obtain its angular spatial frequency spectrum. In the spatial frequency domain, centered at $(0, 0, k_0)$, with $k_0 = 75.1(1/m)$, we sample 14 values in each of the $k_x$, $k_y$, and $k_z$ directions with an interval of $0.046k_0$, resulting in a total of 2744 modes. The corresponding frequency range is $2873 - 5607$ Hz, with a wave packet width of $\sim 70$ cm in the spatial domain and $\sim 2$ ms in the time domain. These 2744 plane wave modes are generated using the 2D acoustic phased array controlled by the multi-channel sound card. The 3D ST acoustic vortices can be characterized either in the $x - y - z$ space by a time snapshot or the $x - y - t$ spatial-temporal domain at a fixed $z$ position. Here, we adopt the latter characterization method, measuring the time-domain signals using microphones at each point within the $z = 0$ plane (the orange region in Fig. 1(c)), as depicted in Figs. 1(c) and 1(d). We employ a separate pulse signal from the sound card to synchronously trigger each recording session.

The 3D iso-intensity profiles of the ST wave packets carrying longitudinal and transverse OAM, obtained through both theoretical calculations and experimental measurements, are shown in Figs. 2(a) and 2(b), respectively. The topological charges for both vortices are set to 1. The time $t = 0$ corresponds to the moment when the center position of the wave packet reaches $z = 0$. The orange arrows in Figs. 2(a) and 2(b) indicate the direction of the OAM carried by the corresponding wave packets. It can be observed that the experimental measurements are consistent with the theoretical predictions. Due to the spiral phase within the $x - y$ domain, there are a series of phase singularities along the $z$-axis, resulting in a tunnel in the direction of the $z$-axis for the wave packet carrying longitudinal OAM, as shown in Fig. 2(a). Similarly, for the wave packet carrying transverse OAM, the phase twists in the $t - x$ domain and a series of phase singularities along the $y$-axis, leading to a tunnel parallel to the $y$-axis, as depicted in Fig. 2(b). We present theoretical calculations and experimental measurements of the amplitude and phase distributions of cross-sections passing through the centers of the two types of wave packets (the cyan planes in Figs. 2(a) and 2(b)). As shown in Figs. 2(c) and 2(d), both cross-sections exhibit annular amplitude distributions, which arise from the presence of phase singularities at the centers of the cross-sections for both types of vortices, resulting in zero intensity. The phase distributions of the cross-sections of the two types of wave packets are provided in Figs. 2(e) and 2(f), respectively. As illustrated in Fig. 2(e), the wave packet carrying longitudinal OAM exhibits a phase screw dislocation in the $x-y$ plane. In contrast, the wave packet carrying transverse OAM exhibits a characteristic edge phase dislocation in the $t-x$ plane cross-section, as depicted in Fig. 2(f). We demonstrate that 3D ST wave packets carrying longitudinal or transverse OAM can be generated in acoustics, laying the foundation for manipulating the direction of OAM.

### *Direct rotation of ST vortices in 3D space.*

While previous studies on acoustic OAM have primarily focused on longitudinal and transverse OAM, it is essential to recognize that acoustic OAM can exhibit arbitrary orientations. To alter the direction of the OAM, we can directly rotate the wave packet carrying transverse or longitudinal OAM in 3D space. Here, we take the example of rotating a wave packet carrying longitudinal OAM. Considering a 45-degree rotation around the $x$-axis of the acoustic pressure field described in Eq. (1), the rotated wave packet is expressed as:

$$p(x,y,z,t) \propto \left[x + i\left(\frac{\sqrt{2}}{2}y - \frac{\sqrt{2}}{2}\zeta\right)\right] e^{-\frac{x^2+y^2+\zeta^2}{w^2}} e^{ik_0\zeta}. \tag{6}$$

The wave packet carrying longitudinal OAM is rotated without altering its propagation direction along the $z$-axis. The direction of the intrinsic OAM carried by the tilted wave packet can be calculated using Eqs. (5) and (6) as $(0, \frac{\sqrt{2}}{2}, \frac{\sqrt{2}}{2})$, indicating that the OAM direction also rotates 45 degrees around the $x$-axis. The 3D iso-intensity profiles, derived from both theoretical calculations and experimental measurements for the tilted wave packet, are illustrated in Figs. 3(a) and 3(b), including perspective, front, side, and top views. The direction of the OAM after rotation is indicated by orange arrows in Figs. 3(a). From the three views, it is evident that there is a tunnel tilted at a 45-degree angle upwards relative to the $z$-axis. This is because the plane containing the spiral phase rotates from the $x-y$ plane to a plane with a normal vector of $(0, \frac{\sqrt{2}}{2}, \frac{\sqrt{2}}{2})$. Figs. 3(c) and 3(d) present the amplitude distributions in the $t-x$, $t-y$, and $x-y$ cross-sections passing through the center of the wave packet, and Figs. 3(e) and 3(f) illustrate the corresponding phase distributions, based on both theoretical calculations and experimental measurements. Due to the rotation of the wave packet, the previously circular intensity distribution on the $x-y$ plane is now rotated onto a plane with a normal vector of $(0, \frac{\sqrt{2}}{2}, \frac{\sqrt{2}}{2})$. The tilted annular intensity distribution obtained after rotation exhibits an elliptical intensity distribution on both the $t-x$ and $x-y$ cross-sections, while the diagonal line with zero intensity in the $t-y$ cross-section corresponds to the tilted tunnel passing through the center of the wave packet, as illustrated in Figs. 3(c) and 3(d). The phase distribution in the $t-x$ cross-section exhibits an edge phase dislocation, corresponding to the transverse OAM component in the $y$-direction, while the phase distribution in the $x-y$ cross-section manifests a helical phase dislocation, corresponding to the longitudinal OAM component in the $z$-direction, as shown in Figs. 3(e) and 3(f). Here, only an example of a wave packet rotating 45 degrees around the $x$-axis is shown. More generally, by controlling the direction and angle of rotation of the wave packet, the orientation of OAM can be precisely manipulated in the 3D space.

*Intersection of vortices carrying longitudinal and transverse OAM.*

To vary the direction of the OAM, apart from rotating a wave packet carrying OAM in a particular direction, another method involves integrating vortices carrying different types of OAM within a single wave packet. For instance, the intersection of a vortex carrying longitudinal OAM and another carrying transverse OAM within a 3D Gaussian wave packet yields a wave packet carrying tilted OAM. The acoustic pressure distribution of the wave packet generated by the intersection of two types of vortices is illustrated as follows:

$$p(x,y,z,t) \propto [x + i\,sgn(l_1)y]^{|l_1|}[\zeta + i\,sgn(l_2)x]^{|l_2|}e^{-\frac{x^2+y^2+\zeta^2}{w^2}}e^{ik_0\zeta}, \quad (7)$$

where $l_1$ and $l_2$ denote the topological charge of the vortices with spiral phase in the $x-y$ domain and $t-x$ domain, respectively. By adjusting the values of $l_1$ and $l_2$, we can change the orientation of the OAM in the $y-z$ plane. When we set the topological charges of both vortices to 1, the direction of the tilted OAM can be calculated using Eqs. (5) and (7), resulting in $(0, 1, 1)$. Figs. 4(a) and 4(b) illustrate the 3D iso-intensity profiles of the wave packet represented by Eq. (7), obtained through both theoretical calculations and experimental measurements. The wave packet features two tunnels, with the $t$-axis direction tunnel arising from the phase singularities of the vortex in the $x-y$ domain and the $y$-axis direction channel originating from the phase singularities in the $t-x$ domain. Figs. 4(c) and 4(d) illustrate the acoustic pressure distribution in the $t-x$, $t-y$, and $x-y$ cross-sections passing through the center of the wave packet.

Furthermore, we demonstrate the intersection of three vortices carrying OAM in mutually perpendicular directions. By changing the topological charges of the three vortices, the orientation of the OAM can be arbitrarily changed within the $x-y-z$ space. Figs. 5(a) and 5(b) depict the theoretical calculation results and experimental measurements of the iso-intensity profiles of the wave packet combining three vortices, each with a topological charge of 1. The calculated intrinsic OAM carried by the wave packet is $(1, 1, 1)$, as depicted by the orange arrow in Figs. 5(a). The tunnel along the $t$-axis is attributed to a series of singularities induced by the vortex in the $x-y$ domain, whereas the tunnels along the $x$-axis and $y$-axis is caused by singularities brought about by the two vortices in the $t-y$ domain and $t-x$ domain. This demonstrates that we can create complex zero-amplitude tunnel structures through the superposition of 3D ST vortices. The acoustic pressure fields of three cross-sections passing through the center of the wave packet, are depicted in Figs. 5(c) and 5(d). In comparison to the wave packet illustrated in Fig. 4, the wave packet depicted in Fig. 5 introduces an extra vortex in the $t-y$ domain, which induces an additional phase difference of $\pi$ between the regions of $t<0$ and $t>0$ in the $t-x$ cross-section, as well as between the regions of $y>0$ and $y<0$ in the $x-y$ cross-section. It is worth noting that, to achieve more precise control over the orientation of OAM in 3D space, the topological charges of the intersected vortices can take fractional values rather than being restricted to integer values.

## *Discussion*

In summary, we report the generation of ST acoustic vortices with arbitrarily oriented OAM. We first extend ST acoustic vortices from 2D to 3D space, constructing ST wave packets carrying longitudinal and transverse OAM, respectively. The ST wave packet carrying tilted OAM can be obtained through the rotation of vortices in 3D space or the intersection of vortices carrying distinct types of OAM. The direction of OAM can be altered in the first method by adjusting the rotation angle or, in the second method, by modifying the topological charge of different types of vortices. The direction of OAM serves as a new degree of freedom, expanding the capabilities for manipulating acoustic waves, with potential applications across various fields. By leveraging the OAM orientation as a reusable degree of freedom, it is possible to enhance the capacity of acoustic communication. Through the interaction between ST vortices and particles, it is possible to transfer OAM in arbitrary directions to particles, thereby achieving transient and arbitrary-directional particle manipulation. Shaping waves in the spatiotemporal domain with increasingly complex structures offers a promising avenue for research in both fundamental physics and applied sciences. Our work provides a platform for studying 3D acoustic wave fields with novel space-time structures, which will aid in the discovery and study of more exotic acoustic wave physics[53–57].

*Acknowledgments.*

The work is jointly supported by the National Key R&D Program of China (Grants No. 2023YFA1406904 and 2021YFB3801801), the Key R&D Program of Jiangsu Province (Grants No. BK20232048 and BK20232015), and the National Natural Science Foundation of China (Grants No. 52250363 and 52203358).

L. S. and G. H. have contributed equally to this work.

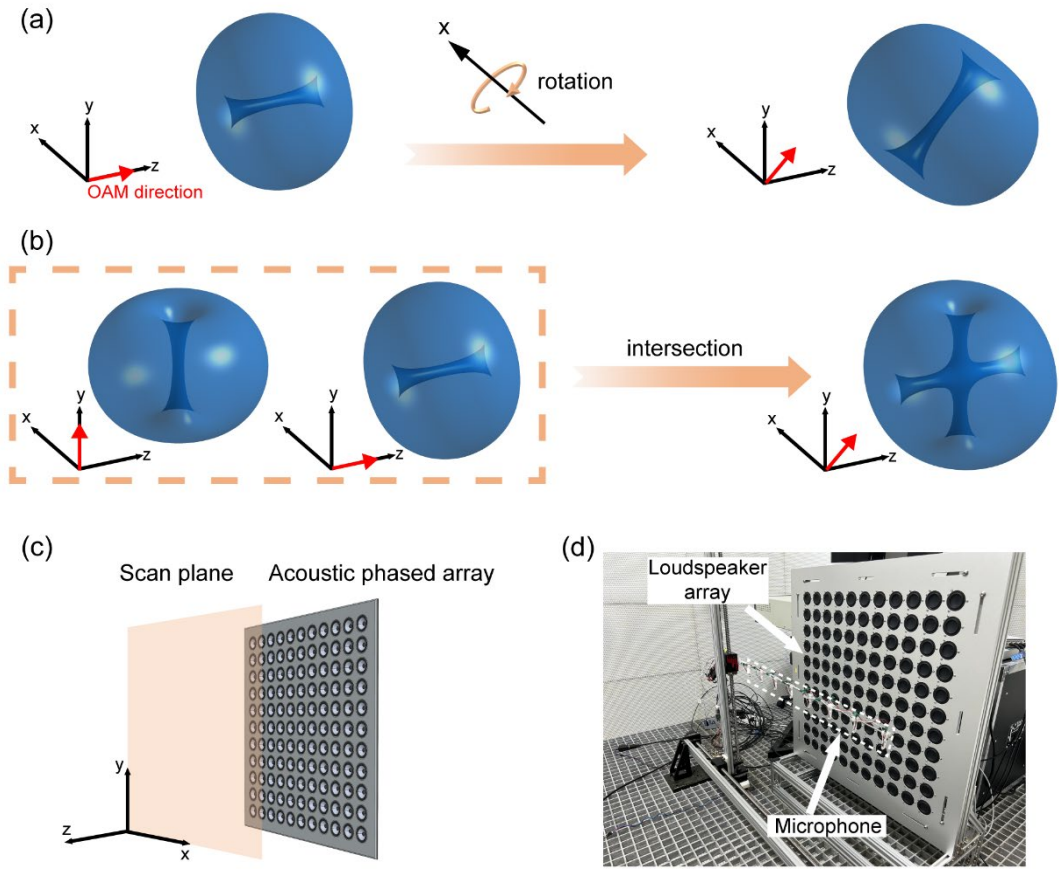

FIG. 1. Synthesizing three-dimensional (3D) spatiotemporal (ST) acoustic vortices with OAM oriented in arbitrary directions. (a) Direct rotation of ST acoustic vortices in 3D space. A ST wave packet carrying longitudinal OAM is rotated around the $x$-axis, and the direction of OAM will change accordingly, which is indicated by the red arrow in the coordinate system. (b) Intersection of ST vortices carrying transverse and longitudinal OAM yields a wave packet carrying tilted OAM. (c) Schematic of experimental setup. The orange region represents the measurement plane. A time sequence of acoustic signals is acquired at each point within the plane. (d) Photograph of the experimental setup. The ST acoustic vortices are generated by an acoustic phased array and captured by a microphone array.

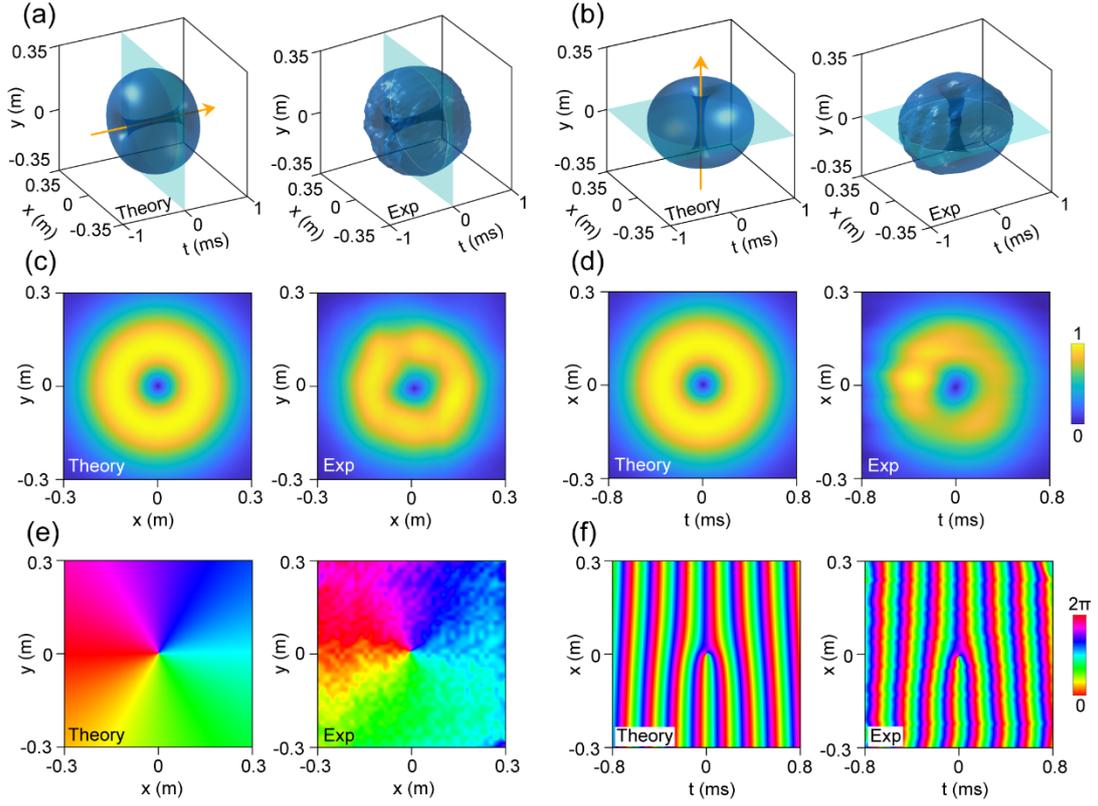

FIG. 2. 3D ST vortices with longitudinal and transverse OAM. (a), (b) The 3D iso-intensity profiles of the ST wave packets carrying longitudinal OAM and transverse OAM. The maximum intensity is set to 1, with isovalue at 0.0625, meaning that the iso-intensity profile includes all points with intensity greater than 0.0625. The orange arrows indicate the direction of OAM carried by the wave packet. (c), (e) The amplitude and phase distributions of the cross-section passing through the center of the wave packets carrying longitudinal OAM (the cyan plane in (a)), which show the screw phase dislocation. (d), (f) The amplitude and phase distributions of the cross-section passing through the center of the wave packets carrying transverse OAM (the cyan plane in (b)), which show the edge phase dislocation.

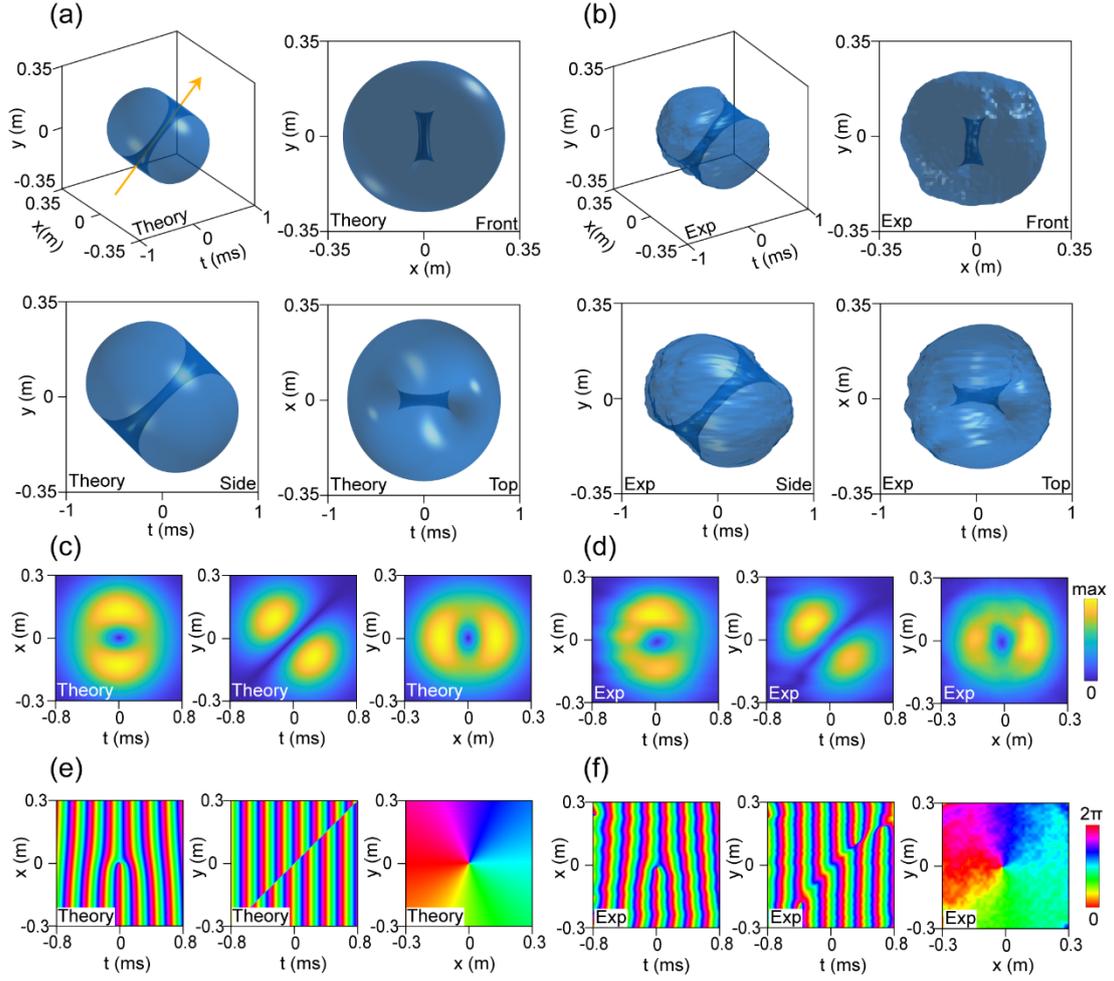

FIG. 3. (a), (b) The 3D iso-intensity profiles of the rotated ST wave packet, along with its front, side, and top views, obtained from theoretical calculations and experimental measurements. The maximum intensity is set to 1, with isovalue at 0.0625. The orange arrows indicate the direction of the OAM carried by the wave packet, pointing towards $(0, \frac{\sqrt{2}}{2}, \frac{\sqrt{2}}{2})$. The amplitude and phase distributions of three cross-sections passing through the center of the wave packet, derived from both theoretical calculations and experimental measurements, are shown in (c), (d) and (e), (f), respectively.

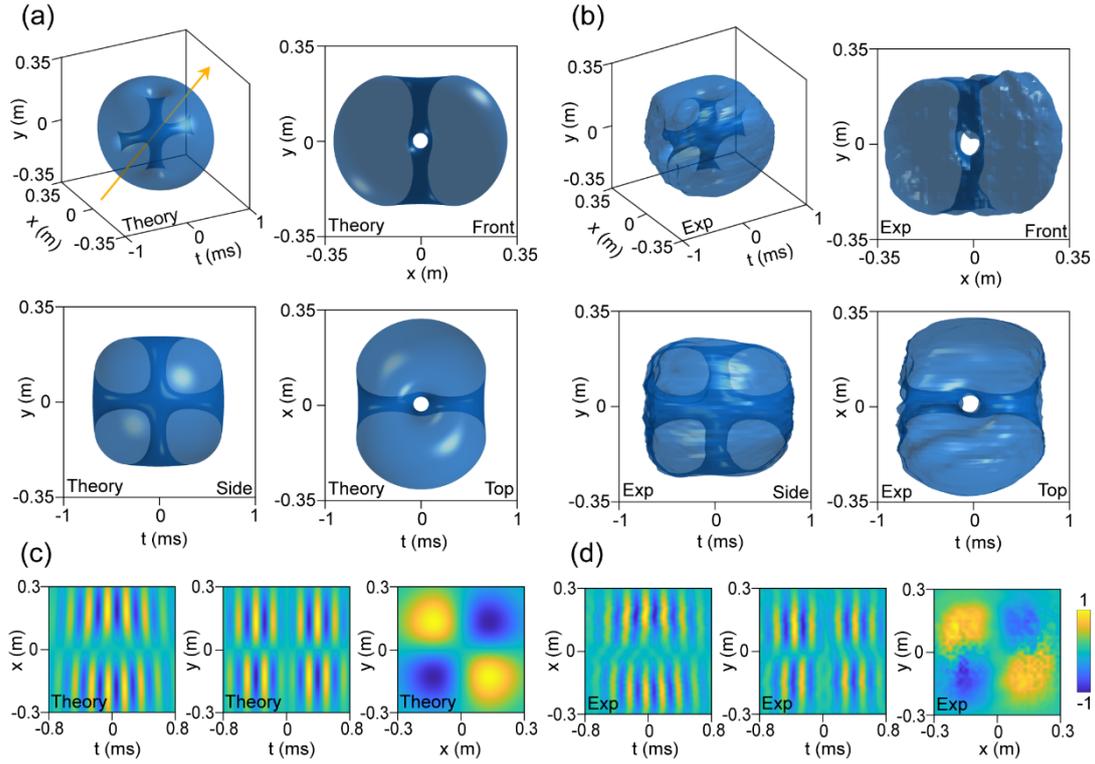

FIG. 4. (a), (b) The 3D iso-intensity profiles of the ST wave packets combining two vortices carrying different types of OAM, obtained through theoretical calculations and experimental measurements. The maximum intensity is set to 1, with isovalue at 0.04. The orange arrows indicate the direction of the OAM carried by the wave packet, pointing towards $(0, 1, 1)$. (c), (d) Theoretical calculation and experimental measurement results of the acoustic pressure field on the cross-sections passing through the center of the wave packet.

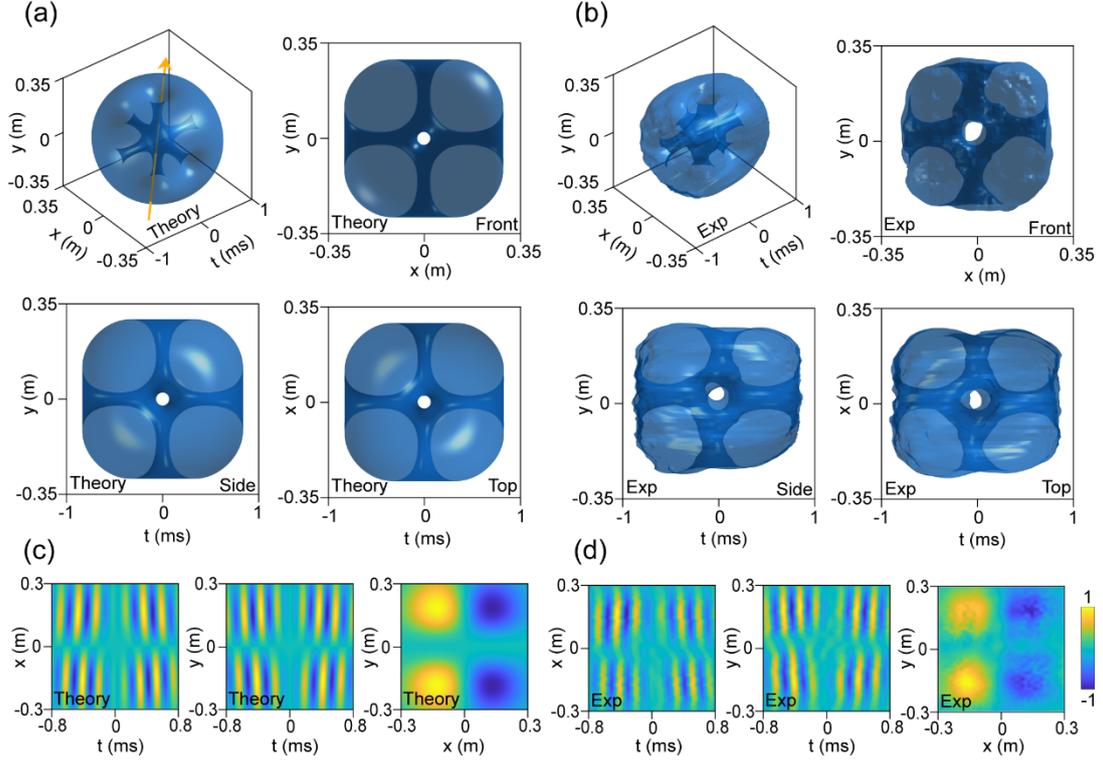

FIG. 5. (a), (b) The 3D iso-intensity profiles of the ST wave packets combining three vortices, obtained through theoretical calculations and experimental measurements. The maximum intensity is set to 1, with isovalue at 0.0841. The orange arrows indicate the direction of the OAM carried by the wave packet, pointing towards $(1, 1, 1)$. (c), (d) Theoretical calculation and experimental measurement results of the acoustic pressure field on the cross-sections passing through the center of the wave packet.